\documentclass{pasa}%

\def\j0621{J0621+2514}
\def\psr{J0621}
\def\fermi{\textit{Fermi}}

\title[Optical identification of the MSP \j0621]{Optical identification of the millisecond pulsar \j0621}
\author[A. V. Karpova et al.]{A. V. Karpova$^1$\thanks{E-mail: annakarpova1989@gmail.com}, 
D. A. Zyuzin$^1$, 
Yu. A. Shibanov$^1$, 
A. Yu. Kirichenko$^{2,1}$
and S. V. Zharikov$^{2}$\\
\affil{$^1$Ioffe Institute, Politekhnicheskaya ul., 26, St. Petersburg, 194021,  Russia}%
\affil{$^2$Instituto de Astronom{\'i}a, Universidad Nacional Aut{\'o}noma de M{\'e}xico, Apdo. Postal 877, Ensenada, Baja California, M{\'e}xico, 22800}
}%

\jid{PASA}
\doi{10.1017/pas.\the\year.xxx}
\jyear{\the\year}

\usepackage{natbib}
\bibpunct{(}{)}{;}{a}{}{,}

\usepackage{pdflscape}
\usepackage{amsmath}

\def\asec{\ifmmode ^{\prime\prime}\else$^{\prime\prime}$\fi}

\def\it{\sl}
\def\degs{\ifmmode ^{\circ}\else$^{\circ}$\fi}
\def\amin{\ifmmode ^{\prime}\else$^{\prime}$\fi}
\def\asec{\ifmmode ^{\prime\prime}\else$^{\prime\prime}$\fi}
\def\fss{\hbox{$.\!\!^{\rm s}$}}        
\def\farcs{\hbox{$.\!\!^{\prime\prime}$}}  

\def\degs{\ifmmode ^{\circ}\else$^{\circ}$\fi}
\def\amin{\ifmmode ^{\prime}\else$^{\prime}$\fi}

\def\eqalign#1{\null\,\vcenter{\openup1\jot \m@th
   \ialign{\strut\hfil$\displaystyle{##}$&$\displaystyle{{}##}$\hfil
   \crcr#1\crcr}}\,}

\begin{document}%

\begin{abstract}
Using the SDSS and Pan-STARRS1 survey data, we found a likely companion of the 
recently discovered binary $\gamma$-ray radio-loud millisecond pulsar \j0621.
Its visual brightness is about 22 mag.
The broad band magnitudes and colours suggest that this is a white dwarf.
Comparing the data with various white dwarfs evolutionary tracks,
we found that it likely belongs to a class of He-core white dwarfs 
with a temperature of about 10~000 K and a mass of $\lesssim 0.5$M$_\odot$.
For a thin hydrogen envelope of the white dwarf its cooling age is $\lesssim0.5$ Gyr
which is smaller than the pulsar characteristic age of 1.8 Gyr.
This may indicate that the pulsar age is overestimated.
Otherwise, this may be explained by the presence of a thick hydrogen envelope 
or a low metallicity of the white dwarf progenitor. 
\end{abstract}
\maketitle%
\section{INTRODUCTION }
\label{sec:intro}

Millisecond pulsars (MSPs) form a special subclass of radio pulsars 
which is characterised by short ($P<30$ ms) rotational periods,
relatively small magnetic fields ($B\sim 10^8-10^{10}$ G)
and low spin-down rates ($\dot{P}\sim 10^{-20}-10^{-19}$ s~s$^{-1}$).
To date about 350 MSPs have been discovered. This number is approximately 13\% of the total number 
of known pulsars\footnote{http://www.atnf.csiro.au/people/pulsar/psrcat/ \citep{atnf2005}}.
It is believed that short spin periods are caused by 
the accretion of matter from a donor star 
\citep{Bisnovatyi-Kogan1974,alpar1982}.
This `recycling' hypothesis is consistent with the fact that most MSPs are found in binary systems. 
The hypothesis is strongly supported by discoveries of transient MSPs 
which show switching between the accretion and radio MSP stages, 
PSR J1023+0038, XSS J12270$-$4859 and PSR J1824$-$2452I 
\citep{Archibald2009,Bassa2014, Roy2015,Papitto2013}.

Observed MSPs companions belong to various stellar classes: main sequence (MS) stars, 
white dwarfs (WDs), non-degenerate or partially degenerate
stellar cores and neutron stars  (NSs) \citep[e.g.][]{manchester2017}.
This depends on initial parameters of the progenitor binary system, e.g.  
the mass of the donor star, the orbital separation,
the environments where the system forms (e.g. the Galactic disc or a globular cluster), etc.
The `fully' recycled  MSPs ($P<10$ ms) most often have He-core WD companions \citep{tauris2011}.

Studies of binary MSPs allow one 
to probe different
astrophysical phenomena including formation and evolution
of binary systems and accretion processes.
The unique rotational stability makes them precise celestial clocks.
This can be used 
to test the accuracy of gravitational theories 
and to search for gravitational waves with a `pulsar timing array' 
\citep{kramer2016,manchester2017}.

Masses of both components in a binary system
can be obtained through a high-precision radio timing by measuring 
the Shapiro delay \citep{shapiro1964}. 
However, this effect is most measurable if an orbit is highly inclined or a companion is massive.
An alternative way to constrain masses and orbital parameters is optical observations \citep[e.g.][]{vankerkwijk2005}. 
In the case of a WD companion, its mass, spectral class and temperature 
can be derived from comparison of photometric and/or
spectroscopic data with predictions of WD cooling models.
The derived WD mass together with the radio timing parameters then can be used 
to estimate the pulsar mass and in turn
to constrain 
the equation of state of the superdense matter in NSs interiors \citep{lattimer2016}.
Optical observations also allow one to obtain the WD cooling age and, therefore,
to constrain the binary system age independent of the pulsar characteristic age.
This is important for studying 
the evolution of binary systems and pulsar spin evolution \citep[e.g.][]{kiziltan2010}. 
  
Binary MSP \j0621\ (hereafter \psr) was recently discovered in radio pulsation searches 
of \fermi\ unassociated sources with the Green Bank Telescope \citep{ray2012,sanpaarsaphd}.
This led to detection of $\gamma$-ray pulsations in the \textit{Fermi} data.
The pulsar parameters obtained from timing analysis are presented in Table~\ref{t:param}.
\citet{sanpaarsaphd} suggested that the system contains a WD companion.
Here we report a likely identification of the pulsar companion using the results of optical surveys.


\begin{figure}
\begin{minipage}[h]{1.\linewidth}
\center{\includegraphics[width=1.0\linewidth,clip]{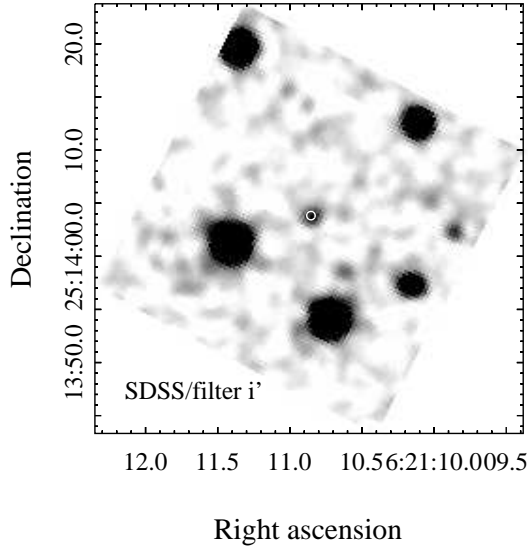}}
\end{minipage}
\caption{SDSS $i'$-band image of the \psr\ field. 
The white circle shows 5$\sigma$ pulsar position uncertainty that accounts 
for the optical astrometric referencing and radio timing position uncertainties from Table~\ref{t:param}.}
\label{fig:1}
\end{figure}

\begin{table}
\caption{Parameters of the \psr\ system obtained from \citet{sanpaarsaphd}.
The distances $D_{\rm YMW}$ and $D_{\rm NE2001}$ are provided from the dispersion measure 
using the YMW16 \citep*{ymw} and NE2001 \citep{ne2001} models for the distribution 
of free electrons in the Galaxy, respectively.} 
\small{
\begin{tabular}{lc}
\hline\hline
Right ascension (RA, J2000)                         & 06$^{\rm h}$21$^{\rm m}$10\fss8542(1) \\
Declination (Dec, J2000)                            & +25\degs14\amin03\farcs83(3)  \\
Spin period $P$ (ms)                                & 2.7217879391872(4) \\
Period derivative $\dot{P}$ ($10^{-21}$ s s$^{-1}$) & 24.83(3) \\
Orbital period $P_b$ (days)                         & 1.256356677(3) \\
Projected semi-major axis (lt-s)                    & 1.276860(3) \\
Dispersion measure (DM, pc cm$^{-3}$)               & 83.629(6) \\
\hline
Characteristic age $\tau_c$(Gyr)                    & 1.8 \\
Surface magnetic field $B$ (G)                      & $2.6\times10^{8}$ \\
Spin-down luminosity $\dot{E}$ (erg s$^{-1}$)       & $4.71\times10^{34}$ \\
Distance $D_{\rm YMW}$ (kpc)                        & 1.64 \\
Distance $D_{\rm NE2001}$ (kpc)                     & 2.33 \\
\hline\hline
\end{tabular}}
 \label{t:param}
\end{table}

\section{The likely companion and its properties} 
\label{sec:companion}

\begin{table*}
\caption{Magnitudes of the \psr\ optical counterpart candidate SDSS J062110.86+251403.8
obtained from SDSS catalogue. }
\begin{center}
\begin{tabular}{lccccccc}
\hline
RA$^\dagger$     & Dec$^\dagger$     & $u'$        & $g'$       & $r'$       & $i'$ & $z'$ \\ %
\hline
06$^{\rm h}$21$^{\rm m}$10\fss861(4)  & +25\degs14\amin03\farcs808(55) & 22.97(37) & 21.90(7) & 21.76(9) &	21.78(13) &	21.23(32) \\
\hline
\end{tabular}
\label{t:psr-coo}
\end{center}
\tabnote{$^\dagger$ Position uncertainties include centroiding and calibration errors 
and calculated as described in \citet{theissen2016} }
\end{table*}

We found  a possible counterpart to \psr\ using the
Sloan Digital Sky Survey Data Release 14 \citep[SDSS DR;][]{sdss14} catalogue.
The position of the point source SDSS J062110.86+251403.8 
overlaps with \psr\ at the 1$\sigma$ significance.
Its position and magnitudes in five filters are presented in Table~\ref{t:psr-coo} and
the $i'$-band image of the pulsar field is shown in Figure~\ref{fig:1}.
The source was also detected in three filters of the Panoramic Survey Telescope and Rapid Response System
Survey \citep[Pan-STARRS;][]{panstarrs1}. 
The probability to detect an unrelated source at the pulsar position is $\sim 10^{-4}$
assuming the source magnitude $r'$ of 18--23.

\begin{figure}
\begin{minipage}[h]{1.\linewidth}
\center{\includegraphics[width=1.0\linewidth,clip]{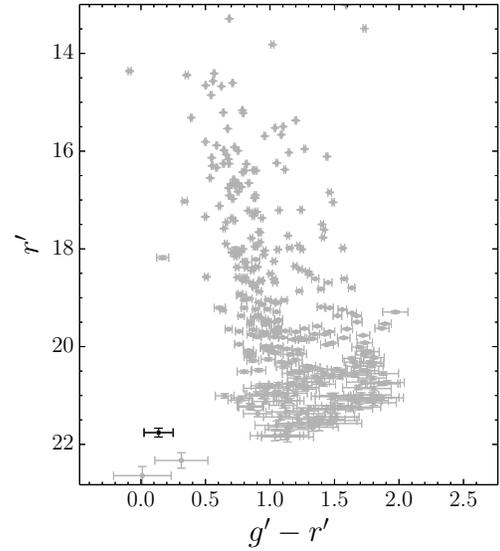}}
\end{minipage}
\caption{Colour-magnitude diagram for sources 
from the SDSS database located within 3 arcmin of the \psr\ position.
The likely pulsar companion is shown in black.}
\label{fig:col-mag_red}
\end{figure}

In Figure~\ref{fig:col-mag_red} we show $g'-r'$ vs $r'$ diagram for sources
from the SDSS database located within 3 arcmin from the \psr\ position. 
One can see that the presumed pulsar companion is shifted bluewards from the MS population 
indicating that it is likely to be a WD.

To obtain dereddened colours of the optical source, 
we estimated the interstellar reddening $E(B-V)$
utilising the dust model by \citet{dustmap2018} 
which is based on the MS stars photometry in the Pan-STARRS~1 and the 2MASS surveys. 
The reddening was then transformed to the extinction correction values using 
conversion coefficients provided by \citet{schlafly2011}. 
For the DM distance $D_{\rm YMW}= 1.64$ kpc we obtained the reddening value 
$E(B-V)=0.25^{+0.02}_{-0.03}$ (see Table~\ref{t:param}).
This implies the dereddened colours  $g^\prime_0-r^\prime_0=-0.12^{+0.14}_{-0.17}$, 
$r^\prime_0-i^\prime_0=-0.17^{+0.17}_{-0.18}$ and 
absolute magnitude $M_{r'} = 10.11^{+0.10}_{-0.11}$
(the errors include uncertainties on the reddening and magnitudes).


\begin{figure}
\begin{minipage}[h]{1.\linewidth}
\center{\includegraphics[width=1.0\linewidth,clip]{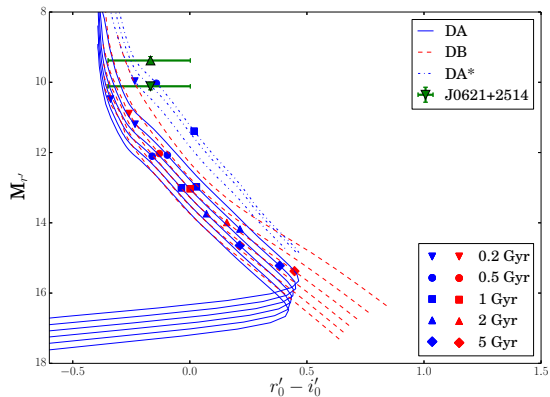}}
\end{minipage}
\caption{Colour-magnitude diagram with various WD cooling tracks. 
Dash-dotted blue lines show tracks for He-core WDs 
with hydrogen atmospheres and masses 0.1869, 0.2026 and 0.2495 ${\rm M_\odot}$ 
\citep[labelled as DA*;][]{panei2007}, solid blue lines --
for WDs with hydrogen atmospheres and masses 0.3--0.8${\rm M_\odot}$ \citep[labelled as DA; the step is 0.1${\rm M_\odot}$;][]{holberg2006,kowalski2006,tremblay2011}, 
red dashed lines -- for WDs with helium atmospheres 
and masses 0.2--0.7${\rm M_\odot}$ \citep[labelled as DB; the step is 0.1${\rm M_\odot}$;][]{bergeron2011}.
WD masses increase from upper to lower curves.
Cooling ages are indicated by different symbols.
The location of the \psr\ companion is marked by the green triangles (the upper one
is for the distance $D_{\rm NE2001}=2.33$ kpc and the lower one -- for $D_{\rm YMW}=1.64$ kpc).}
\label{fig:col-mag}
\end{figure}

\begin{figure}
\begin{minipage}[h]{1.\linewidth}
\center{\includegraphics[width=1.0\linewidth,clip]{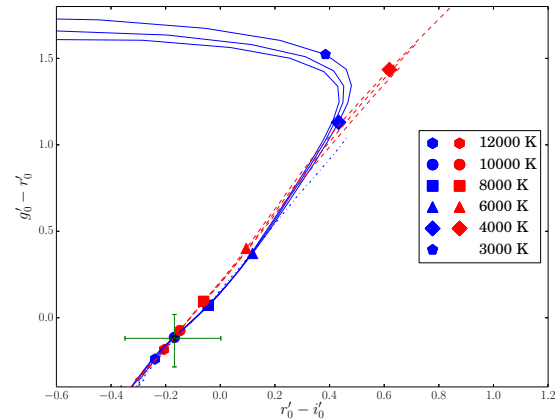}}
\end{minipage}
\caption{Colour-colour diagram with various WD cooling tracks.
The model predictions demonstrated in Fig.~\ref{fig:col-mag} are 
labelled by the same symbols and colours. 
WD temperatures are indicated by different symbols.
The location of the \psr\ companion is marked by the green cross.}
\label{fig:col-col}
\end{figure}

We compared these values with various WD cooling tracks
to check whether the optical source can be indeed a WD.
The colour-magnitude and colour-colour
diagrams are shown in Figures~\ref{fig:col-mag} and \ref{fig:col-col}
where the model predictions for different classes of WDs 
from \citet{panei2007,holberg2006,tremblay2011,kowalski2006,bergeron2011}\footnote{http://www.astro.umontreal.ca/$\sim$bergeron/CoolingModels} are presented.
According to these diagrams, the optical source most likely 
belongs to a class of He-core WDs with 
a mass of $\lesssim 0.5 {\rm M_\odot}$. 
This is a rather young (the cooling age of $\sim$0.2--0.5 Gyr) 
and hot (the effective temperature of $\approx$ $10000\pm2000$ K) WD.
For the distance $D_{\rm NE2001}$=2.3 kpc, 
these estimates remain the same due to large colour uncertainties
(see Fig.~\ref{fig:col-mag}; $M_{r'}=9.38^{+0.10}_{-0.11}$).
 
\section{Discussion and conclusions}
We identified the likely optical counterpart for the binary MSP \psr.
The source is different by its colours from the MS stars (see Figure~\ref{fig:col-mag_red})
which supports its association with the pulsar.
Comparing its absolute magnitudes and colours with WD
cooling tracks, we found that the companion most likely
belongs to the class of the He-core WDs with the temperature $T\approx 10000\pm2000$ K.
The estimated companion mass $M_c$ is less than $0.5 {\rm M_\odot}$. 
We compared the \psr\ period $P$, the orbital period $P_b$
and the companion mass $M_c$ with parameters of other known 
binary MSPs \citep[see, e.g.,][]{manchester2017}.
This system lies in the $M_c$--$P$ and $M_c$--$P_b$ planes 
among others with similar parameters.

From evolutionary tracks we obtain the WD cooling age of $\sim$0.2--0.5 Gyr.
The latter is much shorter than the pulsar characteristic age of 1.8 Gyr.
The situation is similar to PSRs J1012+5307, J1909$-$3744 and J1738+0333 and their WD companions, 
which also have large discrepancies between two ages 
\citep[see, e.g.,][and references therein]{vankerkwijk2005}.
Two explanations of that have been suggested. 
At first, a pulsar age may be overestimated. 
If its initial period is similar to the current one,
then its real age can be compatible with a short cooling age.
\psr\ is a rather energetic pulsar and we cannot exclude that it is young.
Secondly, a WD cooling age may be underestimated. 
WDs can stay hot for a long time due to residual hydrogen burning in the thick hydrogen envelopes.
The latter takes place for the extremely low-mass 
\citep[ELM; $M_{\rm WD}\lesssim 0.2{\rm M_\odot}$;][]{panei2007,vankerkwijk2005} sources. 
Comparison between cooling tracks of ELM WDs with thin and thick hydrogen envelopes was
recently provided by \citet*[][see their Table 1]{calcaferro2018}. 
For example, a WD with $M_{\rm WD}=0.15{\rm M_\odot}$ and thin envelope cools down 
to an effective temperature of $9400$~K in 0.03 Gyr in contrast
with the 2 Gyr required by the thick envelope sequence.
The latter case is in agreement with our results for the \psr+WD system.
The alternative explanation of the WD high temperature is
the low metallicity of its progenitor \citep{serenelli2002}.
CNO flashes in such stars are less intensive than in stars with solar metallicity
and, therefore, even very old WDs remain relatively hot and bright.

At the estimated temperature and mass, spectra of WDs with hydrogen envelopes are 
characterised by a large number of Balmer lines.
Spectroscopic observations may reveal such lines in the spectrum of \psr's putative companion
and allow one to better constrain its temperature, surface gravity, mass and chemical composition.
Radial velocity measurements would confirm the source relation to the pulsar and provide 
the mass ratio and, therefore, the pulsar mass.
For a 22 magnitude source, this is feasible with 8--10 meter class ground-based telescopes.

\begin{acknowledgements}
The authors thank the referee Scott Ransom for the useful comments 
which helped to improve the quality of the paper.
AYuK and SVZ acknowledge PAPIIT grant IN-100617 for resources provided towards this research.
Funding for the Sloan Digital Sky Survey IV has been provided by the Alfred P. Sloan Foundation, 
the U.S. Department of Energy Office of Science, and the Participating Institutions. 
SDSS-IV acknowledges support and resources from the Center for High-Performance Computing at
the University of Utah. The SDSS web site is www.sdss.org.
The Pan-STARRS1 Surveys (PS1) and the PS1 public science archive have been made possible through contributions by the Institute for Astronomy, the University of Hawaii, the Pan-STARRS Project Office, the Max-Planck Society and its participating institutes, the Max Planck Institute for Astronomy, Heidelberg and the Max Planck Institute for Extraterrestrial Physics, Garching, The Johns Hopkins University, Durham University, the University of Edinburgh, the Queen's University Belfast, the Harvard-Smithsonian Center for Astrophysics, the Las Cumbres Observatory Global Telescope Network Incorporated, the National Central University of Taiwan, the Space Telescope Science Institute, the National Aeronautics and Space Administration under Grant No. NNX08AR22G issued through the Planetary Science Division of the NASA Science Mission Directorate, the National Science Foundation Grant No. AST-1238877, the University of Maryland, Eotvos Lorand University (ELTE), the Los Alamos National Laboratory, and the Gordon and Betty Moore Foundation.
\end{acknowledgements}

\bibliographystyle{apj}
\bibliography{msp4}

\begin{thebibliography}{}
\expandafter\ifx\csname natexlab\endcsname\relax\def\natexlab#1{#1}\fi

\bibitem[{{Abolfathi} {et~al.}(2017){Abolfathi}, {Aguado}, {Aguilar}, {Allende
  Prieto}, {Almeida}, {Tasnim Ananna}, {Anders}, {Anderson}, {Andrews},
  {Anguiano}, \& et~al.}]{sdss14}
{Abolfathi}, B., {Aguado}, D.~S., {Aguilar}, G., {et~al.} 2017, ArXiv e-prints,
  arXiv:1707.09322

\bibitem[{{Alpar} {et~al.}(1982){Alpar}, {Cheng}, {Ruderman}, \&
  {Shaham}}]{alpar1982}
{Alpar}, M.~A., {Cheng}, A.~F., {Ruderman}, M.~A., \& {Shaham}, J. 1982,
  Nature, 300, 728

\bibitem[{{Archibald} {et~al.}(2009){Archibald}, {Stairs}, {Ransom}, {Kaspi},
  {Kondratiev}, {Lorimer}, {McLaughlin}, {Boyles}, {Hessels}, {Lynch}, {van
  Leeuwen}, {Roberts}, {Jenet}, {Champion}, {Rosen}, {Barlow}, {Dunlap}, \&
  {Remillard}}]{Archibald2009}
{Archibald}, A.~M., {Stairs}, I.~H., {Ransom}, S.~M., {et~al.} 2009, Science,
  324, 1411

\bibitem[{{Bassa} {et~al.}(2014){Bassa}, {Patruno}, {Hessels}, {Keane},
  {Monard}, {Mahony}, {Bogdanov}, {Corbel}, {Edwards}, {Archibald}, {Janssen},
  {Stappers}, \& {Tendulkar}}]{Bassa2014}
{Bassa}, C.~G., {Patruno}, A., {Hessels}, J.~W.~T., {et~al.} 2014, \mnras, 441,
  1825

\bibitem[{{Bergeron} {et~al.}(2011){Bergeron}, {Wesemael}, {Dufour},
  {Beauchamp}, {Hunter}, {Saffer}, {Gianninas}, {Ruiz}, {Limoges}, {Dufour},
  {Fontaine}, \& {Liebert}}]{bergeron2011}
{Bergeron}, P., {Wesemael}, F., {Dufour}, P., {et~al.} 2011, \apj, 737, 28

\bibitem[{{Bisnovatyi-Kogan} \& {Komberg}(1974)}]{Bisnovatyi-Kogan1974}
{Bisnovatyi-Kogan}, G.~S., \& {Komberg}, B.~V. 1974, Soviet Astronomy, 18, 217

\bibitem[{{Calcaferro} {et~al.}(2018){Calcaferro}, {Althaus}, \&
  {C{\'o}rsico}}]{calcaferro2018}
{Calcaferro}, L.~M., {Althaus}, L.~G., \& {C{\'o}rsico}, A.~H. 2018, ArXiv
  e-prints, arXiv:1802.06753

\bibitem[{{Cordes} \& {Lazio}(2002)}]{ne2001}
{Cordes}, J.~M., \& {Lazio}, T.~J.~W. 2002, ArXiv Astrophysics e-prints,
  astro-ph/0207156

\bibitem[{{Flewelling} {et~al.}(2016){Flewelling}, {Magnier}, {Chambers},
  {Heasley}, {Holmberg}, {Huber}, {Sweeney}, {Waters}, {Chen}, {Farrow},
  {Hasinger}, {Henderson}, {Long}, {Metcalfe}, {Nieto-Santisteban}, {Norberg},
  {Saglia}, {Szalay}, {Rest}, {Thakar}, {Tonry}, {Valenti}, {Werner}, {White},
  {Denneau}, {Draper}, {Hodapp}, {Jedicke}, {Kaiser}, {Kudritzki}, {Price},
  {Wainscoat}, {Chastel}, {McClean}, {Postman}, \& {Shiao}}]{panstarrs1}
{Flewelling}, H.~A., {Magnier}, E.~A., {Chambers}, K.~C., {et~al.} 2016, ArXiv
  e-prints, arXiv:1612.05243

\bibitem[{{Green} {et~al.}(2018){Green}, {Schlafly}, {Finkbeiner}, {Rix},
  {Martin}, {Burgett}, {Draper}, {Flewelling}, {Hodapp}, {Kaiser}, {Kudritzki},
  {Magnier}, {Metcalfe}, {Tonry}, {Wainscoat}, \& {Waters}}]{dustmap2018}
{Green}, G.~M., {Schlafly}, E.~F., {Finkbeiner}, D., {et~al.} 2018, ArXiv
  e-prints, arXiv:1801.03555

\bibitem[{{Holberg} \& {Bergeron}(2006)}]{holberg2006}
{Holberg}, J.~B., \& {Bergeron}, P. 2006, \aj, 132, 1221

\bibitem[{{Kiziltan} \& {Thorsett}(2010)}]{kiziltan2010}
{Kiziltan}, B., \& {Thorsett}, S.~E. 2010, \apj, 715, 335

\bibitem[{{Kowalski} \& {Saumon}(2006)}]{kowalski2006}
{Kowalski}, P.~M., \& {Saumon}, D. 2006, ApJ, 651, L137

\bibitem[{{Kramer}(2016)}]{kramer2016}
{Kramer}, M. 2016, International Journal of Modern Physics D, 25, 1630029

\bibitem[{{Lattimer} \& {Prakash}(2016)}]{lattimer2016}
{Lattimer}, J.~M., \& {Prakash}, M. 2016, \physrep, 621, 127

\bibitem[{{Manchester}(2017)}]{manchester2017}
{Manchester}, R.~N. 2017, Journal of Astrophysics and Astronomy, 38, 42

\bibitem[{{Manchester} {et~al.}(2005){Manchester}, {Hobbs}, {Teoh}, \&
  {Hobbs}}]{atnf2005}
{Manchester}, R.~N., {Hobbs}, G.~B., {Teoh}, A., \& {Hobbs}, M. 2005, \aj, 129,
  1993

\bibitem[{{Panei} {et~al.}(2007){Panei}, {Althaus}, {Chen}, \&
  {Han}}]{panei2007}
{Panei}, J.~A., {Althaus}, L.~G., {Chen}, X., \& {Han}, Z. 2007, \mnras, 382,
  779

\bibitem[{{Papitto} {et~al.}(2013){Papitto}, {Ferrigno}, {Bozzo}, {Rea},
  {Pavan}, {Burderi}, {Burgay}, {Campana}, {di Salvo}, {Falanga},
  {Filipovi{\'c}}, {Freire}, {Hessels}, {Possenti}, {Ransom}, {Riggio},
  {Romano}, {Sarkissian}, {Stairs}, {Stella}, {Torres}, {Wieringa}, \&
  {Wong}}]{Papitto2013}
{Papitto}, A., {Ferrigno}, C., {Bozzo}, E., {et~al.} 2013, Nature, 501, 517

\bibitem[{{Ray} {et~al.}(2012){Ray}, {Abdo}, {Parent}, {Bhattacharya},
  {Bhattacharyya}, {Camilo}, {Cognard}, {Theureau}, {Ferrara}, {Harding},
  {Thompson}, {Freire}, {Guillemot}, {Gupta}, {Roy}, {Hessels}, {Johnston},
  {Keith}, {Shannon}, {Kerr}, {Michelson}, {Romani}, {Kramer}, {McLaughlin},
  {Ransom}, {Roberts}, {Saz Parkinson}, {Ziegler}, {Smith}, {Stappers},
  {Weltevrede}, \& {Wood}}]{ray2012}
{Ray}, P.~S., {Abdo}, A.~A., {Parent}, D., {et~al.} 2012, in Proc. 2011 Fermi
  Symp., eConf C11050

\bibitem[{{Roy} {et~al.}(2015){Roy}, {Ray}, {Bhattacharyya}, {Stappers},
  {Chengalur}, {Deneva}, {Camilo}, {Johnson}, {Wolff}, {Hessels}, {Bassa},
  {Keane}, {Ferrara}, {Harding}, \& {Wood}}]{Roy2015}
{Roy}, J., {Ray}, P.~S., {Bhattacharyya}, B., {et~al.} 2015, \apjl, 800, L12

\bibitem[{{Sanpa-arsa}(2016)}]{sanpaarsaphd}
{Sanpa-arsa}, S. 2016, PhD thesis, University of Virginia

\bibitem[{{Schlafly} \& {Finkbeiner}(2011)}]{schlafly2011}
{Schlafly}, E.~F., \& {Finkbeiner}, D.~P. 2011, \apj, 737, 103

\bibitem[{{Serenelli} {et~al.}(2002){Serenelli}, {Althaus}, {Rohrmann}, \&
  {Benvenuto}}]{serenelli2002}
{Serenelli}, A.~M., {Althaus}, L.~G., {Rohrmann}, R.~D., \& {Benvenuto}, O.~G.
  2002, \mnras, 337, 1091

\bibitem[{{Shapiro}(1964)}]{shapiro1964}
{Shapiro}, I.~I. 1964, Physical Review Letters, 13, 789

\bibitem[{{Tauris}(2011)}]{tauris2011}
{Tauris}, T.~M. 2011, in Astronomical Society of the Pacific Conference Series,
  Vol. 447, Evolution of Compact Binaries, ed. L.~{Schmidtobreick}, M.~R.
  {Schreiber}, \& C.~{Tappert}, 285

\bibitem[{{Theissen} {et~al.}(2016){Theissen}, {West}, \&
  {Dhital}}]{theissen2016}
{Theissen}, C.~A., {West}, A.~A., \& {Dhital}, S. 2016, \aj, 151, 41

\bibitem[{{Tremblay} {et~al.}(2011){Tremblay}, {Bergeron}, \&
  {Gianninas}}]{tremblay2011}
{Tremblay}, P.-E., {Bergeron}, P., \& {Gianninas}, A. 2011, \apj, 730, 128

\bibitem[{{van Kerkwijk} {et~al.}(2005){van Kerkwijk}, {Bassa}, {Jacoby}, \&
  {Jonker}}]{vankerkwijk2005}
{van Kerkwijk}, M.~H., {Bassa}, C.~G., {Jacoby}, B.~A., \& {Jonker}, P.~G.
  2005, in Astronomical Society of the Pacific Conference Series, Vol. 328,
  Binary Radio Pulsars, ed. F.~A. {Rasio} \& I.~H. {Stairs}, 357

\bibitem[{{Yao} {et~al.}(2017){Yao}, {Manchester}, \& {Wang}}]{ymw}
{Yao}, J.~M., {Manchester}, R.~N., \& {Wang}, N. 2017, \apj, 835, 29

\end{thebibliography}

\end{document}